\documentclass[preprint]{aastex631}

\newcommand{\revision}{}

\accepted{\today}

\submitjournal{ApJL}

\shorttitle{ExoFarm}
\shortauthors{Haqq-Misra et al.}
\graphicspath{{./}{figures/}}

\begin{document}

\title{Disruption of a Planetary Nitrogen Cycle as Evidence of Extraterrestrial Agriculture}

\correspondingauthor{Jacob Haqq-Misra}
\email{jacob@bmsis.org}

\author[0000-0003-4346-2611]{Jacob Haqq-Misra}
\affiliation{Blue Marble Space Institute of Science, Seattle, WA, USA}

\author[0000-0002-5967-9631]{Thomas J. Fauchez}
\affiliation{NASA Goddard Space Flight Center, 8800 Greenbelt Road, Greenbelt, MD 20771, USA}
\affiliation{American University, Washington DC, USA}
\affiliation{Sellers Exoplanet Environment Collaboration (SEEC), NASA Goddard Space Flight Center}

\author[0000-0002-2949-2163]{Edward W. Schwieterman}
\affiliation{Department of Earth and Planetary Sciences, University of California, Riverside, CA, USA}
\affiliation{Blue Marble Space Institute of Science, Seattle, WA, USA}

\author[0000-0002-5893-2471]{Ravi Kopparapu}
\affiliation{NASA Goddard Space Flight Center, 8800 Greenbelt Road, Greenbelt, MD 20771, USA}
\affiliation{Sellers Exoplanet Environment Collaboration (SEEC), NASA Goddard Space Flight Center}

\begin{abstract}
Agriculture is one of the oldest forms of technology on Earth. The cultivation of plants requires a terrestrial planet with active hydrological and carbon cycles and depends on the availability of nitrogen in soil. The technological innovation of agriculture is the active management of this nitrogen cycle by applying fertilizer to soil, at first through the production of manure excesses but later by the Haber-Bosch industrial process. The use of such fertilizers has increased the atmospheric abundance of nitrogen-containing species such as NH$_3$ and N$_2$O as agricultural productivity intensifies in many parts of the world. Both NH$_3$ and N$_2$O are effective greenhouse gases, and the combined presence of these gases in the atmosphere of a habitable planet could serve as a remotely detectable spectral signature of technology. Here we use a synthetic spectral generator to assess the detectability of NH$_3$ and N$_2$O that would arise from present-day and future global-scale agriculture. We show that present-day Earth abundances of NH$_3$ and N$_2$O would be difficult to detect but hypothetical scenarios involving a planet with 30-100 billion people could show a change in transmittance of about 50-70\% compared to pre-agricultural Earth. These calculations suggest the possibility of considering the \revision{simultaneous} detection of NH$_3$ and N$_2$O in an atmosphere that also contains H$_2$O, O$_2$, and CO$_2$ as a technosignature for extraterrestrial agriculture. The technology of agriculture is one that could be sustainable across geologic timescales, so the spectral signature of such an ``ExoFarm'' is worth considering in the search for technosignatures.
\end{abstract}

\section{Introduction} \label{sec:intro}

The search for biosignatures seeks to discover evidence of extraterrestrial life through the detection and spectral characterization of exoplanetary atmospheres. Many possibilities for detectable biosignatures have been suggested, \revision{which includes various combinations of CH$_4$, CO$_2$, O$_2$, O$_3$, and H$_2$O based on Earth's history. Specifically,} the concept for searching for \revision{a} combination of \revision{O$_2$ and CH$_4$} gases was first suggested by \citet{lovelock1975thermodynamics} as an example of disequilibrium present in Earth's atmosphere that results from the presence of life. \citet{lovelock1975thermodynamics} observed that the chemical composition of Earth's \revision{present-day} atmosphere remained in a state of thermodynamic disequilibrium, whereas the atmospheric constituents of Venus, Mars, and Jupiter were much closer to an equilibrium state. The combined detection of O$_2$ and CH$_4$ would indicate that a planet has a substantial surface flux of both because CH$_4$ is readily oxidized by O$_2$, and on Earth the major sources of both of these gases are biological. But by themselves, neither O$_2$ nor CH$_4$ would be considered a \revision{compelling} biosignature \citep{schwieterman2018exoplanet}. \revision{Although this example focuses on biosignatures of present-day Earth, similar principles can be applied to ancient Earth \citep[e.g.,][]{arney2016pale,arney2017pale,arney2018organic}.} Additional disequilibria biosignatures that have been suggested include N${_2}$-O${_2}$ and CO${_2}$-CH${_4}$ pairs \citep{krissansen2016detecting, krissansen2018disequilibrium}. \revision{In general, the chemical fluxes and abundances observed in a planet's atmosphere should be evaluated in the context of stellar and planetary characteristics, to assess the potential of a habitable planet to host life.}

The search for technosignatures is a continuation of the search for biosignatures, which includes the idea of looking for spectral evidence of technology in the atmospheres of exoplanets. \revision{The term ``technosignature'' refers generally to any ``evidence of technology that modifies its environment in ways that are detectable'' \citep{2007HiA....14...14T}, which could include a broad class of astronomically observable phenomena. For an overview of modern prospects in the search for technosignatures, see the reviews by \citet{2021AcAau.188..203W}, \citet{2021AcAau.182..446S}, and \citet{lingam2021life}.} A handful of suggestions have been proposed for detectable atmospheric technosignatures, which focus on a single gaseous species as an indicator of extraterrestrial technology. Molecules such as chlorofluorocarbons (CFCs) and halofluorocarbons (HFCs) are examples of industrial products that can have long atmospheric residence times and could be detectable at mid-infrared wavelengths \citep{schneider2010far,lin2014detecting,haqq2022detectability}. Atmospheric pollution could also indicate planetary-scale technology, such as elevated abundances of NO$_2$ due to combustion that could be detectable in the 0.2–0.7 $\mu$m range \citep{kopparapu2021nitrogen}. CFCs and HFCs are almost entirely produced by industry on Earth, while the major sources of NO$_2$ are also industrial, so the detection of these atmospheric constituents in an exoplanetary atmosphere would provide compelling evidence of technology on another planet. 

One of the criticisms of these suggestions is that long-lived technological civilizations may be unlikely to accumulate significant amounts of atmospheric pollution. Industrially-produced constituents such as CFCs and HFCs would only be observable if there were a regular flux into the atmosphere. \revision{One possible scenario could involve the use of industrially-produced greenhouse gases in order to terraform a planet like Mars to make it more habitable \citep{marinova2005radiative,dicaire2013},} but the abundance of such emissions \revision{is restricted} \revision{on Earth today} due to a need to prevent undesirable greenhouse warming by these molecules. Detecting NO$_2$ at elevated abundances would be consistent with a planet engaged in widespread combustion, but combustion itself may not be a sustainable practice over long timescales due to the negative impacts of pollution and finite fuel sources. There may be a large number of atmospheric technosignatures that are unique to industry or cities, but any molecules that are only produced for a short time in the history of a planet---and that do not persist for geologic timescales---will be unlikely to be observed. 

An ideal technosignature would be sustainable for a long time, \revision{as such long-lived evidence would be the most likely to actually be detected \citep{2020IJAsB..19..430K,2021AJ....161..222B}. The two Laser Geodynamics Satellites, known as LAGEOS, are highly reflective satellites used for geodynamics, with no moving parts, that will remain in stable medium-Earth orbits for more than 8 million years \citep{spencer1977lageos}. The LAGEOS satellites are thus an example of a long-duration technosiganture, although the detectability of LAGEOS itself around Earth may be challenging at exoplanetary distances. Another possible long-duration technosignatures is the use of low-albedo energy collectors, which could be detectable by infrared surface imaging \citep{berdyugina2019surface} or spectral signatures in reflected light \citep{lingam2017natural}. Such technosignatures may not be detectable on Earth today, but they represent plausible trajectories for technosignatures in Earth's future.}

An \revision{even more} ideal technosignature would consist of multiple chemical species. This paper suggests that global-scale agriculture provides such a technosignature. \revision{This is not the first time agriculture has been suggested as a technosignature: \citet{1976Icar...28..291S} noted that the possibility of agriculture on Mars could be ruled out based on the lack of checkerboard-like patterns from Mariner 9 imagery. In principle, changes in albedo associated with the timing of crop planting and harvesting could also be detectable by conducting observations over multiple epochs that correspond to different planetary seasons \citep{schwieterman2018exoplanet,2018haex.bookE..69S}, as such changes associated with agriculture can be observed on Earth today \citep{2018NatGe..11...88S}. In this paper, we show that the accumulation of NH$_3$ and N$_2$O from large-scale agriculture is an example of a multi-species and long-lived atmospheric technosignature.}

\section{Agriculture and Nitrogen}

Agriculture is one of the oldest technologies in history. The Agricultural Revolution $\sim$10,000-20,000 years ago began with the end of the last ice age and marked the beginning of permanent human settlements based on agriculture. From a geochemical perspective, agriculture requires a terrestrial planet with an active hydrological cycle as well as a carbon cycle in order to drive photosynthesis. Large-scale photosynthesis could be detectable as an infrared reflectance spectrum that could serve as a biosignature \citep[e.g.][]{kiang2007spectral}, but crop cover alone would be insufficient to serve as a technosignature. Instead, the technological innovation of agriculture is the active management of the nitrogen cycle. The byproducts of this disrupted nitrogen cycle---specifically NH$_3$ and N$_2$O---could serve as atmospheric indicators of agriculture. 

\revision{Nitrogen is an essential nutrient for life, but the vast reservoir of nitrogen in Earth's atmosphere is  unavailable to most organisms because the N$_2$ triple bond is difficult to break. The process of converting N$_2$ into a soluble form is known as nitrogen fixation. Abiotic nitrogen fixation occurs from lightning through the reaction}
\begin{eqnarray}
\mathrm{N_{2}} + \mathrm{2CO_{2}} &\rightarrow& \mathrm{2NO} + \mathrm{2CO},
\label{eq:lightning}
\end{eqnarray}
\revision{which was the only form of nitrogen fixation on pre-biotic Earth and represents about 2\% of total nitrogen fixation today. Biological nitrogen fixation is an anaerobic process that evolved early in the history of life on Earth, which allowed organisms to harvest N$_2$ from the atmosphere. In the oxic environment of Earth today, biological nitrogen fixation is performed by a range of microorganisms using variations on the enzyme nitrogenase in order to break the N$_2$ triple bond, which can be summarized as the reaction}
\begin{eqnarray}
\mathrm{2N_{2}} + \mathrm{6H_{2}O} &\rightarrow& \mathrm{4NH_{3}} + \mathrm{3O_{2}}.
\label{eq:aerobic}
\end{eqnarray}
\revision{In an aqueous environment, the resulting NH$_{3}$ is converted further into the ammonium ion, NH$_{4}^{+}$. These products can be directly taken up by other organisms, or else they can be oxidized into nitrates by nitrifying microorganisms through the reactions}
\begin{eqnarray}
\mathrm{2NH_{4}^{+}} + \mathrm{3O_{2}} &\rightarrow& \mathrm{2NO_{2}^{-}} + \mathrm{2H_{2}O} + \mathrm{4H^{+}}, \\
\label{eq:nitrite}
\mathrm{2NO_{2}^{-}} + \mathrm{O_{2}} &\rightarrow& \mathrm{2NO_{3}^{-}}.
\label{eq:nitrate}
\end{eqnarray}
\revision{The NH$_{4}^{+}$ and NO$_{3}^{-}$ ions provide a form of nitrogen that can be used by other microorganisms, as well as plants, for constructing amino acids. Nitrogen returns to the atmosphere through denitrification, which is an anaerobic processes that mostly occurs in low-oxygen regions of the deep ocean and can be summarized by the reaction}
\begin{eqnarray}
\mathrm{5CH_{2}O} + \mathrm{4NO_{3}^{-}} + \mathrm{4H^{+}} &\rightarrow& \mathrm{5CO_{2}} + \mathrm{2N_{2}} + \mathrm{7H_{2}O}.
\label{eq:return}
\end{eqnarray}
Molecular nitrogen can also return to the atmosphere thorugh the anaerobic anammox pathway ($\mathrm{NH_{4}^{+}} + \mathrm{NO_{2}^{-}} \rightarrow \mathrm{N_{2}} + \mathrm{2H_{2}O}$). \revision{The reactions above describe the nitrogen cycle as it operates on Earth today, and similar processes have been occurring as early as the rise of oxygen $\sim$2.3-2.4\,Gyr ago \citep[e.g.][]{stueken2015isotopic}. For further discussion of Earth's nitrogen cycle, see \citet{sullivan2007planets} and \citet{catling2017atmospheric}.}

Early forms of agriculture relied on manure as the primary source of nitrogen fertilizer, which \revision{was applied directly to fields where the nitrogen would be converted by microorganisms into NH$_{3}$ or NH$_{4}^{+}$ through the process of ammonification. Such practices} increased the demand for animal husbandry and other sources of manure \revision{as populations grew}. Crop rotation was later discovered as a way to replenish soil nitrogen in farmlands by planting nitrogen-fixing crops in alternate years. But the greatest innovation in agriculture, and arguably the most significant discovery of the twentieth century, is the use of the Haber-Bosch process to synthesize ammonia for producing fertilizer. The Haber-Bosch process is a high-temperature industrial process for fixing N$_2$ from the atmosphere \revision{that follows the reaction}
\begin{eqnarray}
\mathrm{N_{2}} + \mathrm{3H_{2}} &\rightarrow& \mathrm{2NH_{3}},
\label{eq:haber}
\end{eqnarray}
\revision{which} uses a metal catalyst \citep{cherkasov2015review}. The main source of H$_2$ today is natural gas, but other sources of H$_2$ such as biomass or water electrolysis also suffice. The ability to manufacture fertilizer using the atmosphere's supply of N$_2$ has allowed farmers to enrich their soils with compounds such as ammonium nitrate (NH$_4$NO$_3$) as a supplement or replacement to urea and manure. These fertilizers release ammonium (NH$_4^+$) and/or nitrate (NO$_3^-$) ions when dissolved in water, which \revision{is then applied to saturate the soil where it can} provide a source of nitrogen to plants. \revision{Excess fertilizer that is not utilized by plant roots contributes to an increase in nitrogen gas emissions, discussed further below.} The Haber-Bosch process revolutionized global agriculture and enabled the production of food surpluses to support a planet populated by billions of people. The expansion of global agriculture has led to an increase in the production of synthetic fertilizers as well as the demand for animal domestication, which leads to an increase in the release of atmospheric nitrogen gases. Indeed, the total anthropogenic fixed nitrogen flux is now equivalent to or greater than non-anthropogenic sources of fixed nitrogen \citep{Battye2017}. 

The most notable \revision{nitrogen-based} atmospheric constituent due to anthropogenic activity is ammonia (NH$_3$). \revision{About 81\% of} the $\sim$58\,Tg\,yr$^{-1}$ of nitrogen in total ammonia emissions \revision{is anthropogenic, with} about 65\% \revision{from} agriculture, 11\% from biomass burning, and 5\% from other industrial processes \citep{seinfeld2016atmospheric}. Only 19\% of NH$_3$ sources are non-anthropogenic, primarily from the volatilization of NH$_3$ from seawater or undisturbed soil as well as from wild animals. The NH$_3$ from agriculture enters the atmosphere from the volatilization of ammonia in soil as well as from domestic animals, all of which originates from fertilizer production \citep{jenkinson2001impact}.  Atmospheric NH$_3$ due to agriculture and animal husbandry has been observed by the Atmospheric Infrared Sounder (AIRS) on the NASA Aqua satellite over a 14-year duration and shows rates of emission that have increased by about 2\% per year, which correlates with increased fertilizer use in some parts of the world \citep{warner2016global,warner2017increased}. The residence time of NH$_3$ in the atmosphere is only hours to days, as most NH$_3$ falls back to the surface through wet or dry deposition. If sufficient NH$_3$ remains in the atmosphere, then it can combine with N$_2$O or SO$_2$ to form aerosol particles. The accumulation of detectable and increasing quantities of NH$_3$ on Earth indicates the intensification of agricultural and industrial activities.

Another significant atmospheric constituent that arises from anthropogenic activity is nitrous oxide (N$_2$O). Of the $\sim$16\,Tg\,yr$^{-1}$ of nitrogen in total N$_2$O emissions, about 40 to 50\% is from agriculture and industry, with the most significant non-anthropogenic sources being the oceans and wet tropical soils \citep{reay2012global, Tian2020}. Most emissions of N$_2$O are the result of denitrification by microorganisms, which in agriculture is enhanced by nitrates added to soil as fertilizer. Other anthropogenic sources of N$_2$O include irrigation water degassing, and animal production---much of which is still ultimately connected to the use of fertilizer---as well as biomass burning. The atmospheric residence time of N$_2$O is about 120\,yr, with the major sink occurring due to photodissociation in the stratosphere with a smaller but significant sink from reactions with O($^{1}$D) radicals. Within the troposphere, N$_2$O is relatively uniformly distributed and also acts as an effective greenhouse gas. The presence of N$_2$O on Earth is generally connected with soil microbiology, but anthropogenic activities that are largely connected with agriculture have enhanced such N$_2$O emissions from soil \citep{Tian2015}. 

Other trace gases are also emitted as the result of agriculture and animal domestication. Agriculture contributes NO and NO$_2$ to the atmosphere from biomass burning and soil denitrification, which accounts for about 25\% of total NO$_x$ emissions \citep{seinfeld2016atmospheric}---although there are large uncertainties with these estimates \citep{jenkinson2001impact}. Nevertheless, a much larger fraction of about 65\% of present-day NO$_x$ emissions are due to fossil fuel combustion; this certainly could serve as a technosignature \citep{kopparapu2021nitrogen}, but NO$_x$ generated from combustion is a separate source from agricultural emissions of NO$_x$ that derive from the application of fertilizer. Methane (CH$_4$) is also emitted from agriculture---notably rice agriculture and ruminant ranching---as well as from biomass burning, landfills, and energy use. About 70\% of total CH$_4$ emissions are anthropogenic, with the rate of these emissions continuing to increase \citep{seinfeld2016atmospheric}. 

Human civilization continues to expand its use of agriculture, and thereby intensify its use of industrial nitrogen fixation to make fertilizer. But there is no particular reason that agriculture itself depends on growth, and as long as sustainable sources of energy are used, then global-scale agriculture could in principle sustain itself across long timescales based on the use of industrial nitrogen fixation \citep[e.g.][]{soloveichik2019electrochemical,smith2020current,wang2021can,rouwenhorst2021beyond}. Whereas processes like combustion may be short-lived due to a finite supply of fossil fuel, the use of industrial nitrogen fixation only requires a planet with a predominantly N$_2$ atmosphere, a supply of H$_2$ and a sustainable source of energy. Thus, the spectral signature of agriculture is well-suited as a candidate for a technosignature that could persist for millennial, and perhaps even geological, timescales. There is little imagination required to speculate that extraterrestrial civilizations, if they exist, would find great value in industrial nitrogen fixation (notably, this is one of the technologies that enables our own civilization to thrive and contemplate our own spectral detectability). 

What, then, is the expected spectral signature of an ``ExoFarm''? The planetary requirements for agriculture as we know it are a hydrological, carbon, and nitrogen cycle, with an atmospheric reservoir of N$_2$ and abundant O$_2$ for photosynthesis. These requirements themselves are aligned with the disequilibrium biosignature of the combined detection of O$_2$ and CH$_4$. In the event that such a planet is discovered, then evidence of elevated levels of NH$_3$ combined with N$_2$O would provide evidence of global-scale agriculture. Because of its extremely short lifetime, the observation of NH$_3$ would imply a continuous large-scale source of emissions, which could be sustained for long periods of time through industrial nitrogen fixation. Although some microorganisms also fix nitrogen, they do not represent significant sources of atmospheric NH$_3$ on Earth. Likewise, the associated detection of N$_2$O and other nitrogen-containing species would provide confidence that the production of NH$_3$ is associated with industrial disruption of a planetary nitrogen cycle.

It is worth emphasizing that NH$_3$ or N$_2$O alone would not necessarily be technosignatures, as either of these species could be false positives for life \citep[e.g.,][]{harman2018biosignature} or could arise from non-technological life \citep[e.g.,][]{roberson2011greenhouse,seager2013biosignature,seager2013biomass,sneed2020climatic,Phillips2021,Huang2022,Ranjan2022}. Rather, it is the combination of NH$_3$ and N$_2$O that would indicate disruption of a planetary nitrogen cycle from an ExoFarm, which may also show elevated abundances of NO$_x$ gases as well as CH$_4$. The short lifetime of NH$_3$ in an oxic atmosphere implies that a detectable abundance of NH$_3$ would suggest a continuous production source. Although NH$_3$ could be produced abiotically by combining N$_2$ and H$_2$, an atmosphere rich in H$_2$ would be unstable to the O$_2$ abundance required to sustain photosynthesis. The technosignature of an ExoFarm would therefore require the simultaneous detection of both NH$_3$ and N$_2$O in the atmosphere of an exoplanet along with O$_2$, H$_2$O, and CO$_2$.

\section{Detectability Constraints}\label{sec:detect}

Large-scale agriculture based on Haber-Bosch nitrogen fixation could be detectable through the infrared spectral absorption features of NH$_3$ and N$_2$O as well as CH$_4$. A robust assessment of the detectability of such spectral features in an Earth-like atmosphere would ideally use a three-dimensional coupled climate-chemistry model to calculate the steady-state abundances of each of these nitrogen-containing species a function of biological and technological surface fluxes. But as an initial assessment, we first consider a scaling argument to examine the spectral features that could be detectable for present-day and future Earth agriculture.

We define four scenarios for considering agriculture on an Earth-like planet, with the corresponding atmospheric abundances of nitrogen-containing species listed in Table \ref{tab:scenarios}. The present-day Earth scenario is based on recent measurements of NH$_3$, N$_2$O, and CH$_4$ abundances \citep{seinfeld2016atmospheric,ipcc6th}. The choice of 10\,ppb for NH$_3$ is toward the higher-end for Earth today and corresponds to regions of intense agricultural production. 

\begin{table}[ht!]
\centering
\caption{Agricultural scenarios with estimated atmospheric abundances of nitrogen-containing species.\label{tab:scenarios}}
\begin{tabular}{cccc}
\hline
Scenario                & NH$_3$ (ppb) & N$_2$O (ppb) & CH$_4$ (ppb) \\
\hline\hline
Pre-agricultural Earth  & 2     & 170   & 570   \\
Present-day Earth       & 10    & 335   & 1900   \\
Future Earth 30B        & 30    & 590   & 4300   \\
Future Earth 100B       & 100   & 1900  & 14000  \\
\hline                       
\end{tabular}
\end{table}

The pre-agricultural Earth scenario serves as a control, where the agricultural and technological contributions of NH$_3$, N$_2$O, and CH$_4$ have been removed. Note that this approach assumes that eliminating the technological contributions to the atmospheric flux of these nitrogen-containing species will reduce the steady-state atmospheric abundance by a similar percentage; this approach is admittedly simplified, but the results can still be instructive for identifying the possibility of detectable spectral features. 

The third and fourth scenarios project possible abundances of NH$_3$, N$_2$O, and CH$_4$ for futures with 30 and 100 billion people, respectively. Earth holds about 7.9 billion people today, and population projections differ on whether or not Earth's population will stabilize in the coming century \citep{gerland2014world,warren2015can,vollset2020fertility}. These two population values were selected because they correspond approximately to the maximum total allowable population using all current arable land ($\sim$ 30 billion) and all possible agricultural land ($\sim$ 100 billion) \citep{mullan2019population}. \revision{Most published estimates of Earth's carrying capacity range from about 8 to 100 billion, although some estimates are less than 1 billion while others are more than 1 trillion \citep{cohen1995population}. Theoretically, an extraterrestrial population with the energy requirements of up to 100 billion calorie-consuming humans could sustain Haber-Bosch synthesis over long timescales, as long as sustainable energy sources are used \citep[e.g.][]{soloveichik2019electrochemical,smith2020current,wang2021can,rouwenhorst2021beyond}.} These scenarios also follow a scaling argument by assuming that the per person contributions of these three nitrogen-containing species will remain constant as population grows. This again is a simplifying assumption that is intended as an initial approach to understanding the detectability of such scenarios.

\begin{figure}[ht!]
\centering
\includegraphics[width=0.75\linewidth]{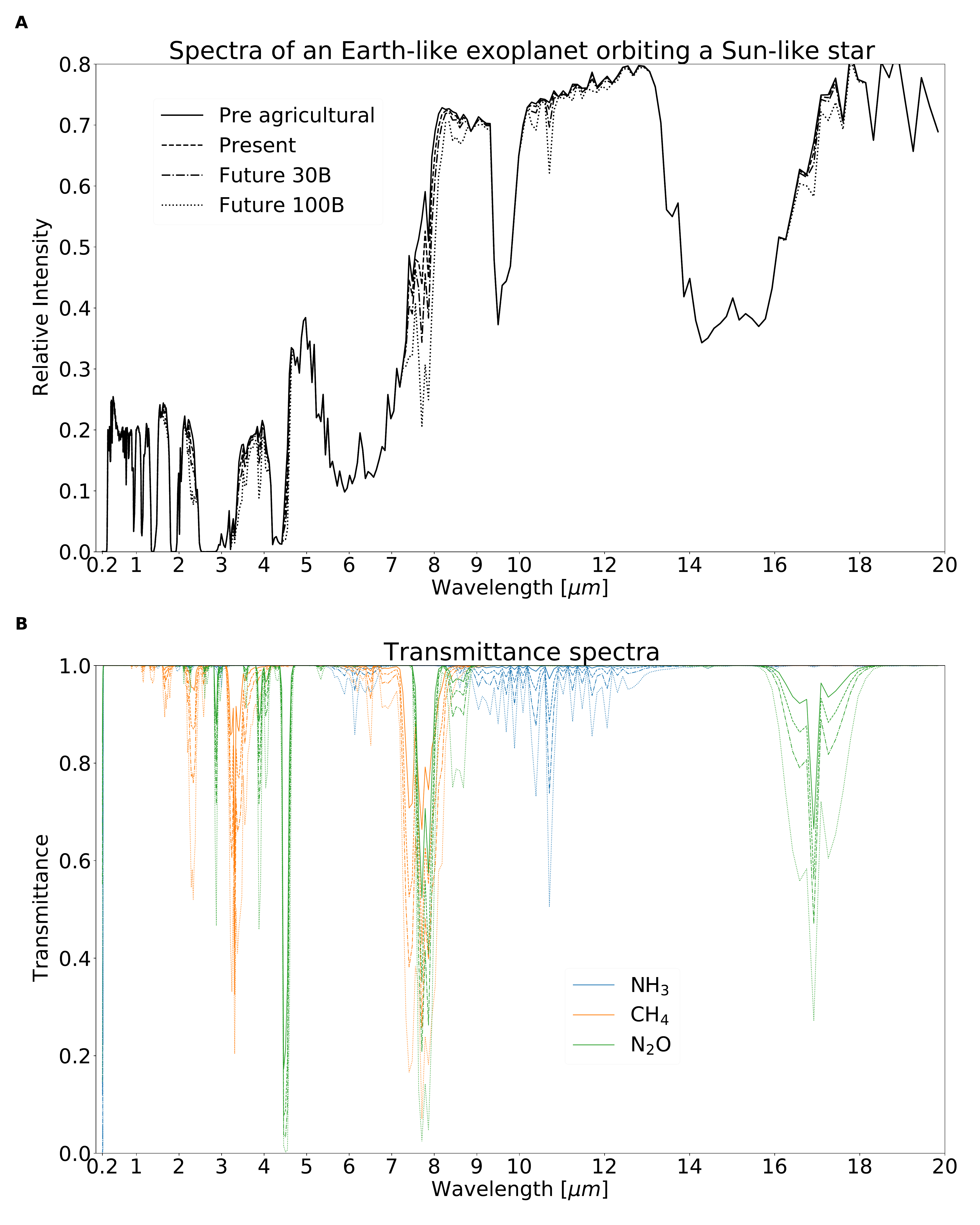}
\caption{Infrared spectral features of NH$_3$, N$_2$O, and CH$_4$ for pre-agricultural, present-day, and future Earth scenarios, with relative intensity shown in the top panel and transmittance shown in the bottom panel. Calculations are performed with the Planetary Spectrum generator.
\label{fig:spectra}}
\end{figure}

We consider the detectability of all four of these scenarios using the Planetary Spectrum Generator (PSG \citet{villanueva2018planetary,PSG2022}). PSG is an online radiative transfer tool for calculating synthetic planetary spectra and assessing the limits of detectability for spectral features that can range from ultraviolet to radio wavelengths. The ultraviolet features of NH$_3$, N$_2$O, and CH$_4$ are strongly overlapping and only show weak absorption, but mid-infrared features of all these species could be more \revision{pronounced}. The mid-infrared spectral features of NH$_3$, N$_2$O, and CH$_4$ calculated with PSG for pre-agricultural, present-day, and future Earth scenarios are plotted in Fig. \ref{fig:spectra}, which shows the relative intensity (top) and transmittance spectra (bottom) for observations of an Earth-like exoplanet orbiting a sun-like star.

The spectra shown in Fig. \ref{fig:spectra} show the strongest absorption features due to NH$_3$ from 10-12\,$\mu$m, while N$_2$O shows absorption features from 3-5\,$\mu$m, 7-9\,$\mu$m, and 16-18\,$\mu$m. Absorption features due to CH$_4$ overlap some of the N$_2$O features from 3-5\,$\mu$m and 7-9\,$\mu$m. The change in \revision{peak transmittance between 10-12\,$\mu$m (bottom panel, Fig. \ref{fig:spectra})} for NH$_3$ compared to the pre-agricultural control case is about 50\% for the future Earth scenario with 100 billion people and about 25\% for the scenario with 30 billion people. For N$_2$O, the change in \revision{peak transmittance between 16-18\,$\mu$m compared to the pre-agricultural control case} is about 70\% for 100 billion people and 50\% for 30 billion people. The change in relative intensity (top panel, Fig. \ref{fig:spectra} for the 100 billion people scenario is up to about 10\% \revision{compared to the pre-agricultural control case} between 7-9\,$\mu$m and 10-12\,$\mu$m. Present-day Earth agriculture would exert a weakly detectable signal that might be difficult to discern from the pre-agricultural control case, but future scenarios with enhanced global agriculture could produce absorption features that are easier to detect.

The spectral features of NH$_3$, N$_2$O, and CH$_4$ could be detectable in emitted light or as transmission features for transiting planets. \revision{Specifically, the N$_2$O line at 17.0\,$\mu$m shows a strong dependency with the N$_2$O volume mixing ratio and to a second order the NH$_3$ line at 10.7\,$\mu$m. For the Future 100B case, both display strong enough absorption to be detectable by the LIFE (Large Interferometer for Exoplanets, \citet{quanz2021atmospheric,2021arXiv210107500L}), Origins \citep{meixner2019origins}, and MIRECLE (Mid-InfraRed Exo-planet CLimate Explorer, \citet{staguhn2019mid}) infrared mission concepts. The James Webb Space Telescope (JWST) Near Infrared Spectrograph (NIRSpec) could potentially detect CH$_4$ within the 0.6-5.3\,$\mu$m range for transiting exoplanets \citep{krissansen2018disequilibrium}. However, the detection of CH$_4$ alone would provide no basis for distinguishing between technological, biological, or photochemical production. The detectability of these spectral features do not necessarily directly correspond to the peak transmittance, and a full accounting of the detectability of each band would need to account for the observing mode and instrument parameters.} It is beyond the scope of this present paper to present detectability calculations for specific missions, as any missions capable of searching for mid-infrared technosignatures are in an early design phase, at best. One of the goals of this paper is to highlight the importance of examining mid-infrared spectral features of exoplanets, as many potential technosignatures could be most detectable at such wavelengths. Also, it demonstrates the duality of the search for biosignatures and technosignatures. The search for passive, atmospheric technosignatures does not require the development of a dedicated instrument but can leverage the capability of instruments dedicated to the search for biosignature.

Another, more technically challenging possibility for detecting industrial N-fixation would be constraining the $^{15}$N to $^{14}$N isotopic ratios of N-bearing atmospheric gases, including NH$_{3}$ and N$_{2}$O. The Haber-Bosch process introduces a well-known depletion in $\delta^{15}$N, the “$^{15}$N Haber-Bosch effect" \citep{Yang2016}. Future work would be required to assess the spectral detectability of industrial $^{15}$N/$^{14}$N impacts, but would certainly require high resolution spectroscopy. 

We note that other studies have considered the role of NH$_3$ as a biosignature and its possible detectability. \citet{Huang2022} examined the accumulation of NH$_3$ in an optimal environment of a hydrogen-rich planet orbiting an M-dwarf star and concluded that the James Webb Space Telescope could detect such spectral features at NH$_3$ abundances of about 5\,ppm. This is more than an order of magnitude greater than the NH$_3$ abundance in our future earth scenario with 100 billion people. Even so, such speculation by others suggests a need for broader thinking with regard to the possible biosignatures and technosignatures that could be detectable.

\section{Next Steps}

The calculations presented in this paper indicate the possibility of detecting a technosignature from planetary-scale agriculture from the combined the spectral features of NH$_3$ and N$_2$O, as well as CH$_4$. The signature of such an ExoFarm could only occur on a planet that already supports photosynthesis, so such a planet will necessarily already show spectral features due to H$_2$O, O$_2$, and CO$_2$. The search for technosignatures from extraterrestrial agriculture would therefore be a goal that supports the search for biosignatures of Earth-like planets, as the best targets to search for signs of nitrogen cycle disruption would be planets already thought to be good candidates for photosynthetic life.

A better constraint on the detectability of the spectral features of an ExoFarm would require the use of an atmospheric photochemistry model. This paper assumed simple scaling arguments for the abundances of nitrogen-containing species, but the steady-state abundance of nitrogen-containing atmospheric species will depend on a complex network of chemical reactions \revision{and the photochemical impact of the host star's UV spectrum}. \revision{In such future work, the increases of NH$_3$ and N$_2$O, and CH$_4$ from agriculture would be parameterized via surface fluxes instead of arbitrary fixed and vertically constant mixing ratios. A network of photochemical reactions would then determine the vertical distribution of those species in the atmosphere.} A photochemical model could also capture the processes of wet and dry deposition of NH$_3$, \revision{which is the major sink in Earth's present atmosphere}, as well as aerosol formation from NH$_3$ and SO$_2$/N$_2$O that can occur in regions of high agricultural production. \revision{Past studies have predicted more favorable build-up of biosignature gases on oxygen-rich Earth-like planets orbiting later spectral type (K- or M-type) stars due to orders of magnitude less efficient production of OH, O($^1$D), and other radicals that attack trace gases like CH$_4$ \citep{segura2005biosignatures,arney2019k}. The photochemical lifetime of N$_2$O and therefore its steady state mixing ratio will be enhanced by less efficient production of O($^1$D) radicals that destroy it. However, because deposition is the major sink of NH$_{3}$, it is not clear whether a different stellar environment would alter the atmospheric lifetime of NH$_{3}$, and if so, to what extent. The application of an appropriate photochemical model could answer this unknown.} 

Examining the four scenarios in this study with such a photochemical model would require additional development work to extend the capabilities of existing models to oxygen-rich atmospheres. Past photochemical modeling studies that have included NH$_{3}$ considered anoxic Early Earth scenarios where the focus was determining the plausible greenhouse impact of NH$_{3}$ to revolve the Faint Young Sun Paradox \citep{kasting1982stability,pavlov2001uv}. More recent studies have considered NH$_{3}$ biosignatures in H$_{2}$-dominated super-Earth atmospheres, which would greatly favor the spectral detectability of the gas relative to high molecular weight O$_{2}$-rich atmospheres \citep{seager2013biosignature,seager2013biomass,sneed2020climatic,Phillips2021}. \revision{On H$_2$ planets with surfaces saturated with NH$_3$, deposition is inefficient, and sufficient biological fluxes can overwhelm photochemical sinks and can allow large NH$_3$ mixing ratios to be maintained \citep{Huang2022, Ranjan2022}. These ``Cold Haber Worlds'' are far different from the O$_2$-N$_2$ atmosphere we consider here, where surfaces saturated in NH$_3$ are implausible and photochemical lifetimes are shorter.} Ideally, future calculations would use a three-dimensional model with coupled climate and photochemical processes suitable for an O$_2$-N$_2$ atmosphere to more completely constrain the steady-state abundances, and time variation, in nitrogen-containing species for planets with intensive agriculture.

\revision{Future investigation should also consider false positive scenarios for NH$_3$ and N$_2$O as a technosignature. One possibility is that a species engages in global-scale agriculture using manure only; such a planet could conceivably accumulate detectable quantities of NH$_3$ and N$_2$O without the use of the Haber-Bosch process. The distinction between these two scenarios might be difficult to resolve, but both forms of agriculture nevertheless represent a technological innovation. Whether or not similar quantities of NH$_3$ and N$_2$O could accumulate on a planet by animal-like life without active management is a possible area for future work. External factors such as stellar proton events associated with flares could also produce high abundances of nitrogen-contaning species in an atmosphere rich in NH$_3$ \citep[e.g.][]{2017NatSR...714141A}, so additional false positive scenarios should be considered for planets in systems with high stellar activity.}

This paper is intended to present the idea that the spectral signature of extraterrestrial technology would be a compelling technosignature. This does not necessarily imply that extraterrestrial agriculture must necessarily exist or be commonplace, but the idea of searching for spectral features of an ExoFarm remains a plausible technosignature based on future projections of Earth today. Such a technosignature could also be long-lived, perhaps on geologic timescales, and would indicate the presence of a technological species that has managed to co-exist with technology while avoiding extinction. Long-lived technosignatures are the most likely to be discovered by astronomical means, so scientists engaged in the search for technosignatures should continue to think critically about technological processes that could be managed across geologic timescales.

\begin{acknowledgments}
J.H.M. gratefully acknowledges support from the NASA Exobiology program under grant 80NSSC20K0622. E.W.S. acknowledges support from the NASA ICAR program. T.J.F and R.K.K. acknowledges support from the GSFC Sellers Exoplanet Environments Collaboration (SEEC), which is supported by NASA's Planetary Science Divisions Research Program. Any opinions, findings, and conclusions or recommendations expressed in this material are those of the authors and do not necessarily reflect the views of their employers or NASA.
\end{acknowledgments}

\bibliography{main}{}

\begin{thebibliography}{}
\expandafter\ifx\csname natexlab\endcsname\relax\def\natexlab#1{#1}\fi
\providecommand{\url}[1]{\href{#1}{#1}}
\providecommand{\dodoi}[1]{doi:~\href{http://doi.org/#1}{\nolinkurl{#1}}}
\providecommand{\doeprint}[1]{\href{http://ascl.net/#1}{\nolinkurl{http://ascl.net/#1}}}
\providecommand{\doarXiv}[1]{\href{https://arxiv.org/abs/#1}{\nolinkurl{https://arxiv.org/abs/#1}}}

\bibitem[{{Airapetian} {et~al.}(2017){Airapetian}, {Jackman}, {Mlynczak},
  {Danchi}, \& {Hunt}}]{2017NatSR...714141A}
{Airapetian}, V.~S., {Jackman}, C.~H., {Mlynczak}, M., {Danchi}, W., \& {Hunt},
  L. 2017, Scientific Reports, 7, 14141, \dodoi{10.1038/s41598-017-14192-4}

\bibitem[{Arney {et~al.}(2018)Arney, Domagal-Goldman, \&
  Meadows}]{arney2018organic}
Arney, G., Domagal-Goldman, S.~D., \& Meadows, V.~S. 2018, Astrobiology, 18,
  311

\bibitem[{Arney {et~al.}(2016)Arney, Domagal-Goldman, Meadows, Wolf,
  Schwieterman, Charnay, Claire, H{\'e}brard, \& Trainer}]{arney2016pale}
Arney, G., Domagal-Goldman, S.~D., Meadows, V.~S., {et~al.} 2016, Astrobiology,
  16, 873

\bibitem[{Arney(2019)}]{arney2019k}
Arney, G.~N. 2019, The Astrophysical Journal Letters, 873, L7,
  \dodoi{10.3847/2041-8213/ab065}

\bibitem[{Arney {et~al.}(2017)Arney, Meadows, Domagal-Goldman, Deming,
  Robinson, Tovar, Wolf, \& Schwieterman}]{arney2017pale}
Arney, G.~N., Meadows, V.~S., Domagal-Goldman, S.~D., {et~al.} 2017, The
  Astrophysical Journal, 836, 49

\bibitem[{{Balbi} \& {{\'C}irkovi{\'c}}(2021)}]{2021AJ....161..222B}
{Balbi}, A., \& {{\'C}irkovi{\'c}}, M.~M. 2021, \aj, 161, 222,
  \dodoi{10.3847/1538-3881/abec48}

\bibitem[{Battye {et~al.}(2017)Battye, Aneja, \& Schlesinger}]{Battye2017}
Battye, W., Aneja, V.~P., \& Schlesinger, W.~H. 2017, Earth's Future, 5, 894,
  \dodoi{10.1002/2017EF000592}

\bibitem[{Berdyugina \& Kuhn(2019)}]{berdyugina2019surface}
Berdyugina, S., \& Kuhn, J. 2019, The Astronomical Journal, 158, 246

\bibitem[{Catling \& Kasting(2017)}]{catling2017atmospheric}
Catling, D.~C., \& Kasting, J.~F. 2017, Atmospheric evolution on inhabited and
  lifeless worlds (Cambridge University Press)

\bibitem[{Cherkasov {et~al.}(2015)Cherkasov, Ibhadon, \&
  Fitzpatrick}]{cherkasov2015review}
Cherkasov, N., Ibhadon, A., \& Fitzpatrick, P. 2015, Chemical Engineering and
  Processing: Process Intensification, 90, 24

\bibitem[{Cohen(1995)}]{cohen1995population}
Cohen, J.~E. 1995, Science, 269, 341

\bibitem[{Dicaire {et~al.}(2013)Dicaire, Forget, Millour, Maan, Nachon, \&
  Summerer}]{dicaire2013}
Dicaire, I., Forget, F., Millour, E., {et~al.} 2013, in 64th International
  Astronautical Congress, IAC-13 (Beijing, China: IAF), D3, 3.10x19180

\bibitem[{Gerland {et~al.}(2014)Gerland, Raftery,
  {\v{S}}ev{\v{c}}{\'\i}kov{\'a}, Li, Gu, Spoorenberg, Alkema, Fosdick, Chunn,
  Lalic, {et~al.}}]{gerland2014world}
Gerland, P., Raftery, A.~E., {\v{S}}ev{\v{c}}{\'\i}kov{\'a}, H., {et~al.} 2014,
  Science, 346, 234

\bibitem[{Haqq-Misra {et~al.}(2022)Haqq-Misra, Kopparapu, Fauchez, Frank,
  Wright, \& Lingam}]{haqq2022detectability}
Haqq-Misra, J., Kopparapu, R., Fauchez, T.~J., {et~al.} 2022, The Planetary
  Science Journal, 3, 60

\bibitem[{Harman \& Domagal-Goldman(2018)}]{harman2018biosignature}
Harman, C.~E., \& Domagal-Goldman, S. 2018, Handbook of Exoplanets, 71

\bibitem[{Huang {et~al.}(2022)Huang, Seager, Petkowski, Ranjan, \&
  Zhan}]{Huang2022}
Huang, J., Seager, S., Petkowski, J.~J., Ranjan, S., \& Zhan, Z. 2022,
  Astrobiology, 22, 171, \dodoi{10.1089/ast.2020.2358}

\bibitem[{IPCC(2021)}]{ipcc6th}
IPCC. 2021, IPCC, Geneva, 2021

\bibitem[{Jenkinson(2001)}]{jenkinson2001impact}
Jenkinson, D.~S. 2001, Plant and Soil, 228, 3

\bibitem[{Kasting(1982)}]{kasting1982stability}
Kasting, J.~F. 1982, Journal of Geophysical Research: Oceans, 87, 3091

\bibitem[{Kiang {et~al.}(2007)Kiang, Siefert, \&
  Blankenship}]{kiang2007spectral}
Kiang, N.~Y., Siefert, J., \& Blankenship, R.~E. 2007, Astrobiology, 7, 222

\bibitem[{{Kipping} {et~al.}(2020){Kipping}, {Frank}, \&
  {Scharf}}]{2020IJAsB..19..430K}
{Kipping}, D., {Frank}, A., \& {Scharf}, C. 2020, International Journal of
  Astrobiology, 19, 430, \dodoi{10.1017/S1473550420000208}

\bibitem[{Kopparapu {et~al.}(2021)Kopparapu, Arney, Haqq-Misra, Lustig-Yaeger,
  \& Villanueva}]{kopparapu2021nitrogen}
Kopparapu, R., Arney, G., Haqq-Misra, J., Lustig-Yaeger, J., \& Villanueva, G.
  2021, The Astrophysical Journal, 908, 164

\bibitem[{Krissansen-Totton {et~al.}(2016)Krissansen-Totton, Bergsman, \&
  Catling}]{krissansen2016detecting}
Krissansen-Totton, J., Bergsman, D.~S., \& Catling, D.~C. 2016, Astrobiology,
  16, 39

\bibitem[{Krissansen-Totton {et~al.}(2018)Krissansen-Totton, Olson, \&
  Catling}]{krissansen2018disequilibrium}
Krissansen-Totton, J., Olson, S., \& Catling, D.~C. 2018, Science advances, 4,
  eaao5747

\bibitem[{{LIFE collaboration} {et~al.}(2021){LIFE collaboration}, {Quanz},
  {Ottiger}, {Fontanet}, {Kammerer}, {Menti}, {Dannert}, {Gheorghe}, {Absil},
  {Airapetian}, {Alei}, {Allart}, {Angerhausen}, {Blumenthal}, {Cabrera},
  {Carri{\'o}n-Gonz{\'a}lez}, {Chauvin}, {Danchi}, {Dandumont}, {Defr{\`e}re},
  {Dorn}, {Ehrenreich}, {Ertel}, {Fridlund}, {Garc{\'\i}a Mu{\~n}oz},
  {Gasc{\'o}n}, {Glauser}, {Grenfell}, {Guidi}, {Hagelberg}, {Helled},
  {Ireland}, {Kopparapu}, {Korth}, {Kraus}, {L{\'e}ger}, {Leedj{\"a}rv},
  {Lichtenberg}, {Lillo-Box}, {Linz}, {Liseau}, {Loicq}, {Mahendra}, {Malbet},
  {Mathew}, {Mennesson}, {Meyer}, {Mishra}, {Molaverdikhani}, {Noack}, {Oza},
  {Pall{\'e}}, {Parviainen}, {Quirrenbach}, {Rauer}, {Ribas}, {Rice},
  {Romagnolo}, {Rugheimer}, {Schwieterman}, {Serabyn}, {Sharma}, {Stassun},
  {Szul{\'a}gyi}, {Wang}, {Wunderlich}, \& {Wyatt}}]{2021arXiv210107500L}
{LIFE collaboration}, {Quanz}, S.~P., {Ottiger}, M., {et~al.} 2021, arXiv
  e-prints, arXiv:2101.07500.
\newblock \doarXiv{2101.07500}

\bibitem[{Lin {et~al.}(2014)Lin, Abad, \& Loeb}]{lin2014detecting}
Lin, H.~W., Abad, G.~G., \& Loeb, A. 2014, The Astrophysical Journal Letters,
  792, L7

\bibitem[{Lingam \& Loeb(2017)}]{lingam2017natural}
Lingam, M., \& Loeb, A. 2017, Monthly Notices of the Royal Astronomical
  Society: Letters, 470, L82

\bibitem[{Lingam \& Loeb(2021)}]{lingam2021life}
---. 2021, Life in the Cosmos: From Biosignatures to Technosignatures (Harvard
  University Press)

\bibitem[{Lovelock(1975)}]{lovelock1975thermodynamics}
Lovelock, J.~E. 1975, Proceedings of the Royal Society of London. Series B.
  Biological Sciences, 189, 167

\bibitem[{Marinova {et~al.}(2005)Marinova, McKay, \&
  Hashimoto}]{marinova2005radiative}
Marinova, M.~M., McKay, C.~P., \& Hashimoto, H. 2005, Journal of Geophysical
  Research: Planets, 110

\bibitem[{Meixner {et~al.}(2019)Meixner, Cooray, Leisawitz, Staguhn, Armus,
  Battersby, Bauer, Bergin, Bradford, Ennico-Smith,
  {et~al.}}]{meixner2019origins}
Meixner, M., Cooray, A., Leisawitz, D., {et~al.} 2019, arXiv preprint
  arXiv:1912.06213

\bibitem[{Mullan \& Haqq-Misra(2019)}]{mullan2019population}
Mullan, B., \& Haqq-Misra, J. 2019, Futures, 106, 4

\bibitem[{Pavlov {et~al.}(2001)Pavlov, Brown, \& Kasting}]{pavlov2001uv}
Pavlov, A.~A., Brown, L.~L., \& Kasting, J.~F. 2001, Journal of Geophysical
  Research: Planets, 106, 23267

\bibitem[{Phillips {et~al.}(2021)Phillips, Wang, Kendrew, Greene, Hu, Valenti,
  Panero, \& Schulze}]{Phillips2021}
Phillips, C.~L., Wang, J., Kendrew, S., {et~al.} 2021, The Astrophysical
  Journal, 923, 144, \dodoi{10.3847/1538-4357/ac29be}

\bibitem[{Quanz {et~al.}(2021)Quanz, Absil, Benz, Bonfils, Berger, Defr{\`e}re,
  van Dishoeck, Ehrenreich, Fortney, Glauser, {et~al.}}]{quanz2021atmospheric}
Quanz, S.~P., Absil, O., Benz, W., {et~al.} 2021, Experimental Astronomy, 1

\bibitem[{Ranjan {et~al.}(2022)Ranjan, Seager, Zhan, Koll, Bains, Petkowski,
  Huang, \& Lin}]{Ranjan2022}
Ranjan, S., Seager, S., Zhan, Z., {et~al.} 2022.
\newblock \doarXiv{2201.08359}

\bibitem[{Reay {et~al.}(2012)Reay, Davidson, Smith, Smith, Melillo, Dentener,
  \& Crutzen}]{reay2012global}
Reay, D.~S., Davidson, E.~A., Smith, K.~A., {et~al.} 2012, Nature climate
  change, 2, 410

\bibitem[{Roberson {et~al.}(2011)Roberson, Roadt, Halevy, \&
  Kasting}]{roberson2011greenhouse}
Roberson, A.~L., Roadt, J., Halevy, I., \& Kasting, J. 2011, Geobiology, 9, 313

\bibitem[{Rouwenhorst {et~al.}(2021)Rouwenhorst, Van~der Ham, \&
  Lefferts}]{rouwenhorst2021beyond}
Rouwenhorst, K.~H., Van~der Ham, A.~G., \& Lefferts, L. 2021, international
  journal of hydrogen energy, 46, 21566

\bibitem[{{Sagan} \& {Lederberg}(1976)}]{1976Icar...28..291S}
{Sagan}, C., \& {Lederberg}, J. 1976, \icarus, 28, 291,
  \dodoi{10.1016/0019-1035(76)90039-7}

\bibitem[{Schneider {et~al.}(2010)Schneider, L{\'e}ger, Fridlund, White, Eiroa,
  Henning, Herbst, Lammer, Liseau, Paresce, {et~al.}}]{schneider2010far}
Schneider, J., L{\'e}ger, A., Fridlund, M., {et~al.} 2010, Astrobiology, 10,
  121

\bibitem[{{Schwieterman}(2018)}]{2018haex.bookE..69S}
{Schwieterman}, E.~W. 2018, in Handbook of Exoplanets, ed. H.~J. {Deeg} \&
  J.~A. {Belmonte}, 69, \dodoi{10.1007/978-3-319-55333-7\_69}

\bibitem[{Schwieterman {et~al.}(2018)Schwieterman, Kiang, Parenteau, Harman,
  DasSarma, Fisher, Arney, Hartnett, Reinhard, Olson,
  {et~al.}}]{schwieterman2018exoplanet}
Schwieterman, E.~W., Kiang, N.~Y., Parenteau, M.~N., {et~al.} 2018,
  Astrobiology, 18, 663

\bibitem[{Seager {et~al.}(2013{\natexlab{a}})Seager, Bains, \&
  Hu}]{seager2013biosignature}
Seager, S., Bains, W., \& Hu, R. 2013{\natexlab{a}}, The Astrophysical Journal,
  777, 95

\bibitem[{Seager {et~al.}(2013{\natexlab{b}})Seager, Bains, \&
  Hu}]{seager2013biomass}
---. 2013{\natexlab{b}}, The Astrophysical Journal, 775, 104

\bibitem[{Segura {et~al.}(2005)Segura, Kasting, Meadows, Cohen, Scalo, Crisp,
  Butler, \& Tinetti}]{segura2005biosignatures}
Segura, A., Kasting, J.~F., Meadows, V., {et~al.} 2005, Astrobiology, 5, 706,
  \dodoi{10.1089/ast.2005.5.706}

\bibitem[{Seinfeld \& Pandis(2016)}]{seinfeld2016atmospheric}
Seinfeld, J.~H., \& Pandis, S.~N. 2016, Atmospheric chemistry and physics: from
  air pollution to climate change (John Wiley \& Sons)

\bibitem[{{Seneviratne} {et~al.}(2018){Seneviratne}, {Phipps}, {Pitman},
  {Hirsch}, {Davin}, {Donat}, {Hirschi}, {Lenton}, {Wilhelm}, \&
  {Kravitz}}]{2018NatGe..11...88S}
{Seneviratne}, S.~I., {Phipps}, S.~J., {Pitman}, A.~J., {et~al.} 2018, Nature
  Geoscience, 11, 88, \dodoi{10.1038/s41561-017-0057-5}

\bibitem[{Smith {et~al.}(2020)Smith, Hill, \&
  Torrente-Murciano}]{smith2020current}
Smith, C., Hill, A.~K., \& Torrente-Murciano, L. 2020, Energy \& Environmental
  Science, 13, 331

\bibitem[{Sneed(2020)}]{sneed2020climatic}
Sneed, E.~L. 2020, A Climatic Investigation of Ammonia as a Remote Biosignature
  on Cold Haber Worlds, \dodoi{10.26207/9c0r-j916}

\bibitem[{{Socas-Navarro} {et~al.}(2021){Socas-Navarro}, {Haqq-Misra},
  {Wright}, {Kopparapu}, {Benford}, {Davis}, \& {TechnoClimes 2020 workshop
  participants}}]{2021AcAau.182..446S}
{Socas-Navarro}, H., {Haqq-Misra}, J., {Wright}, J.~T., {et~al.} 2021, Acta
  Astronautica, 182, 446, \dodoi{10.1016/j.actaastro.2021.02.029}

\bibitem[{Soloveichik(2019)}]{soloveichik2019electrochemical}
Soloveichik, G. 2019, Nature Catalysis, 2, 377

\bibitem[{Spencer(1977)}]{spencer1977lageos}
Spencer, R.~L. 1977, Journal of Geological Education, 25, 38

\bibitem[{Staguhn {et~al.}(2019)Staguhn, Mandell, Stevenson, Saxena, Kopparapu,
  Fixsen, Sharp, DiPirro, Knez, Wolf, {et~al.}}]{staguhn2019mid}
Staguhn, J., Mandell, A., Stevenson, K., {et~al.} 2019, arXiv preprint
  arXiv:1908.02356

\bibitem[{St{\"u}eken {et~al.}(2015)St{\"u}eken, Buick, Guy, \&
  Koehler}]{stueken2015isotopic}
St{\"u}eken, E.~E., Buick, R., Guy, B.~M., \& Koehler, M.~C. 2015, Nature, 520,
  666

\bibitem[{Sullivan \& Baross(2007)}]{sullivan2007planets}
Sullivan, W.~T., \& Baross, J. 2007, Planets and life: the emerging science of
  astrobiology (Cambridge University Press)

\bibitem[{{Tarter}(2007)}]{2007HiA....14...14T}
{Tarter}, J.~C. 2007, Highlights of Astronomy, 14, 14,
  \dodoi{10.1017/S1743921307009829}

\bibitem[{Tian {et~al.}(2015)Tian, Chen, Lu, Xu, Ren, Zhang, Banger, Tao, Pan,
  Liu, Zhang, Bruhwiler, \& Wofsy}]{Tian2015}
Tian, H., Chen, G., Lu, C., {et~al.} 2015, Ecosystem Health and Sustainability,
  1, 1, \dodoi{10.1890/EHS14-0015.1}

\bibitem[{Tian {et~al.}(2020)Tian, Xu, Canadell, Thompson, Winiwarter,
  Suntharalingam, Davidson, Ciais, Jackson, Janssens-Maenhout, Prather,
  Regnier, Pan, Pan, Peters, Shi, Tubiello, Zaehle, Zhou, Arneth, Battaglia,
  Berthet, Bopp, Bouwman, Buitenhuis, Chang, Chipperfield, Dangal, Dlugokencky,
  Elkins, Eyre, Fu, Hall, Ito, Joos, Krummel, Landolfi, Laruelle, Lauerwald,
  Li, Lienert, Maavara, MacLeod, Millet, Olin, Patra, Prinn, Raymond, Ruiz,
  van~der Werf, Vuichard, Wang, Weiss, Wells, Wilson, Yang, \& Yao}]{Tian2020}
Tian, H., Xu, R., Canadell, J.~G., {et~al.} 2020, Nature, 586, 248,
  \dodoi{10.1038/s41586-020-2780-0}

\bibitem[{{Villanueva} {et~al.}(2022){Villanueva}, {Liuzzi}, {Faggi},
  {Protopapa}, {Kofman}, {Stone}, \& {Mandell}}]{PSG2022}
{Villanueva}, G.~L., {Liuzzi}, G., {Faggi}, S., {et~al.} 2022, {Fundamentals of
  the Planetary Spectrum Generator}

\bibitem[{Villanueva {et~al.}(2018)Villanueva, Smith, Protopapa, Faggi, \&
  Mandell}]{villanueva2018planetary}
Villanueva, G.~L., Smith, M.~D., Protopapa, S., Faggi, S., \& Mandell, A.~M.
  2018, Journal of Quantitative Spectroscopy and Radiative Transfer, 217, 86

\bibitem[{Vollset {et~al.}(2020)Vollset, Goren, Yuan, Cao, Smith, Hsiao,
  Bisignano, Azhar, Castro, Chalek, {et~al.}}]{vollset2020fertility}
Vollset, S.~E., Goren, E., Yuan, C.-W., {et~al.} 2020, The Lancet, 396, 1285

\bibitem[{Wang {et~al.}(2021)Wang, Khan, Mohsin, Wicks, Ip, Sumon, Dinh,
  Sargent, Gates, \& Kibria}]{wang2021can}
Wang, M., Khan, M.~A., Mohsin, I., {et~al.} 2021, Energy \& Environmental
  Science, 14, 2535

\bibitem[{Warner {et~al.}(2017)Warner, Dickerson, Wei, Strow, Wang, \&
  Liang}]{warner2017increased}
Warner, J., Dickerson, R., Wei, Z., {et~al.} 2017, Geophysical Research
  Letters, 44, 2875

\bibitem[{Warner {et~al.}(2016)Warner, Wei, Strow, Dickerson, \&
  Nowak}]{warner2016global}
Warner, J.~X., Wei, Z., Strow, L.~L., Dickerson, R.~R., \& Nowak, J.~B. 2016,
  Atmospheric Chemistry and Physics, 16, 5467

\bibitem[{Warren(2015)}]{warren2015can}
Warren, S.~G. 2015, Earth's Future, 3, 82

\bibitem[{{Wright}(2021)}]{2021AcAau.188..203W}
{Wright}, J.~T. 2021, Acta Astronautica, 188, 203,
  \dodoi{10.1016/j.actaastro.2021.07.021}

\bibitem[{Yang \& Gruber(2016)}]{Yang2016}
Yang, S., \& Gruber, N. 2016, Global Biogeochemical Cycles, 30, 1418,
  \dodoi{10.1002/2016GB005421}

\end{thebibliography}
\bibliographystyle{aasjournal}

\end{document}